# Microstructure and structural modulation of lutetium dihydride LuH$_2$ as seen via transmission electron microscopy

Xiao-Ping Ma,[1, 2, 3, ‡] Ning-Ning Wang,[1, 5, ‡] Wen-Tao Wang,[1, 5] Jing-Zhe Nie,[1, 5] Wen-Li Gao,[1] Shuai-Shuai Sun,[1, 5] Jun Li,[1, 5] Huan-Fang Tian,[1, 5] Tian-Long Xia,[2, 3, 4, 6] Jin-Guang Cheng,[1, 5, *] Jian-Qi Li,[1, 5, 7, *] and Huai-Xin Yang[1, 5, *]

1 Beijing National Laboratory for Condensed Matter Physics and Institute of Physics, Chinese Academy of Sciences, Beijing 100190, China

2 Department of Physics, Renmin University of China, Beijing 100872, China

3 Beijing Key Laboratory of Opto-electronic Functional Materials & Micro-nano Devices, Renmin University of China, Beijing 100872, China

4 Key Laboratory of Quantum State Construction and Manipulation (Ministry of Education), Renmin University of China, Beijing, 100872, China

5 School of Physical Sciences, University of Chinese Academy of Sciences, Beijing 100190, China

6 Laboratory for Neutron Scattering, Renmin University of China, Beijing 100872, China

7 Songshan Lake Materials Laboratory, Dongguan, Guangdong, 523808, China

**ABSTRACT:** Structural investigations conducted using transmission electron microscopy (TEM) on LuH$_2$ synthesized under atmospheric pressure (AP-LuH$_2$) and nitrogen-doped LuH$_2$ synthesized under high pressure (HP-LuH$_2$) have revealed numerous microstructural phenomena. Both materials show a clear superstructure modulation with wave vector, $\boldsymbol{q}^* = \frac{1}{4}(2\bar{2}0)$, and this modulation can be well interpreted by the displacements of Lu atoms. Further investigations on the nitrogen-doped HP-LuH$_2$ materials reveal the appearance of high-density antiphase boundaries, in particular, domain walls of a few atomic layer thickness without structural modulation can be observed, suggesting possible interface properties could be detected in this system. *In-situ* TEM observations of AP-LuH$_2$ suggest that no evident structural phase transition occurs between 94 K and 673 K.

## ■ INTRODUCTION

Metallic hydrogen is expected to display a high Debye temperature and strong electron-phonon coupling, which could lead to high-temperature superconductivity (SC) based on the Bardeen-Cooper–Schrieffer (BCS) theory[1]. However, achieving the hydrogen metallization is difficult owing to the extremely high metallized pressure[2]. The polyhydride approach has been proposed because of its chemical precompression effect[3, 4], followed by the theoretical prediction of sulfur hydrides, SH$_2$ and SH$_3$, to host high-temperature SC with critical temperature of approximately $Tc$ = 80 and 204 K at 160 and 200 GPa, respectively[5, 6]. Furthermore, high-temperature SC was experimentally observed with $Tc$ ≈ 203 K under 155 GPa in a sulfur hydride system[7]. Following the discovery of sulfur hydride superconductor, lanthanide polyhydrides were synthesized and found to be superconducting with $Tc$ of 250-260 K at 170-200 GPa for LaH$_{10}$[8-11], 243-262 K at 180-201 GPa for YH$_9$[12, 13], ~220 K at 183 GPa for YH$_6$[13], and $Tc$ of ~210 K at 160-172 GPa for CaH$_6$[14, 15].

For the lanthanide polyhydrides, SC was observed to relate with the 4$f$ electrons and $Tc$ was found to decrease with increasing 4$f$ electrons because of the spin scattering effects, LaH$_{10}$ displayed the highest $Tc$ of 260 K, whereas the maximum $Tc$ of CeH$_{10}$ decreased to 115 K[16], further decreasing to 9 and 5 K for PrH$_9$[17] and NdH$_9$[18], respectively. These experimental observations conform with the theoretical predictions about the $f$ electrons dependence on $Tc$ for the early lanthanide polyhydrides[19]. However, for lutetium lanthanide with fully filled $f$ orbitals, the contribution of the 4$f$ electrons to the electric density-of-state near the Fermi level must be minimal, and its effect on the SC of the polyhydride should be minimized. Lutetium and lanthanum have similar electronegativity and ability to provide electrons to dissociate hydrogen molecules to atoms. Thus, lutetium polyhydride is expected to host the highest $Tc$ SC due to its fully filled 4$f$ shell. Theoretical studies on the Lu-H system have predicted high-temperature superconductivity at relatively low pressures, e.g., the $Im$-$3m$ LuH$_6$ with $Tc$ = 273 K at 100 GPa[20]. Shao *et al.* experimentally obtained a superconducting LuH$_3$ with the critical temperature $Tc$ in the range of 12−15 K over the pressure range of 110–170 GPa[21]. Li *et al.* reported that when the synthesis pressure of the above-mentioned reaction in the diamond anvil cell (DAC) is elevated to 184 GPa, a new polyhydride is formed with the possible formula of Lu$_4$H$_{23}$, which adopts a higher H content and shows higher $Tc$ reaching 65–71 K over 181–218 GPa[22].

Recently, Dias *et al.* had reported room-temperature SC of 294 K at a remarkably low pressure of 1 GPa in an Lu-N-H system[23]. Based on the X-ray diffraction (XRD) results, the superconducting phase is attributed to LuH$_{3-δ}$N$_ε$. Additionally, the sample's color was observed to change from blue to pink at the pressure of 0.3–3 GPa (phase II) and then to red at pressures above 3 GPa (phase III)[23]. Moreover, the pink-colored phase

II was reported to exhibit room-temperature SC with the maximum $T_c$ of ~294 K at approximately 1 GPa. These discoveries attracted research attention at a global scale in terms of Lu hydrides. However, several different groups have reported no evidence of near-ambient SC reproduction in the Lu-N-H system, synthesized using the same method following Dias's report or other alternative methods[24, 25]. Hemley et al. obtained samples synthesized by the Dias group and independently characterized the electrical transport measurements in DACs[26]. They observed evidence of SC on the nitrogen-doped lutetium hydride with a maximum $T_c$ of 276 K at 15 kbar, which is consistent with previous results by Dias et al. [23]. The SC was suggested to be strongly dependent on material synthesis conditions, including the synthesis method, nitrogen and hydrogen sources, the concentration of nitrogen doping, and the annealing process [26]. These results indicate that the successful synthesis of the superconducting material is strongly dependent on the details of sample preparation. Although the transport and magnetic properties of $LuH_2$ or $LuH_{2+x}N_y$ have been investigated under high pressure, their microstructure, especially on the atomic scale, has not been focused on much.

Scanning transmission electron microscopy (STEM) technique with sub-angstrom spatial resolution has been extensively applied for structural characterizations, especially for structural modulation, phase separation, and local defective structures. In this paper, a systematic microscopic characterization of $LuH_2$ synthesized under atmospheric pressure (AP-$LuH_2$) and nitrogen-doped $LuH_2$ synthesized under high pressure (HP-$LuH_2$) have performed using STEM. The experimental results demonstrate that a superstructure phase with wave vector, $q^* = \frac{1}{4}(2\bar{2}0)$, originating from the ordering displacement of Lu, can be identified in the parent compound, $LuH_2$, and high density of antiphase boundaries in nitrogen-doped HP-$LuH_2$. These HP-$LuH_2$ materials display domain walls of a few atomic layer thickness without structural modulation.

### ■ RESULT AND DISCUSSION

The polycrystalline sample of $LuH_2$ (AP-$LuH_2$) material was synthesized with temperature 200 °C for 5 hours using solid-state method, as described in a previous paper[27]. The nitrogen doped HP-$LuH_2$ samples were prepared by sintering a mixture of AP-$LuH_2$ and LuN powders in the ratio of 95:5 under high-pressure (10 GPa) and high-temperature (600°C) conditions in a Kawai-type multianvil module (Max Voggenreiter GmbH). The complete synthetic protocols can be found in the Supporting Information.

Lu dihydride $LuH_2$ with fluorite structure showed stability at ambient conditions, displaying metallic conductivity[28, 29]. Figure 1(a) shows the structural model of $LuH_2$, in which Lu and H atoms form a face-centered cubic and octahedral frameworks, respectively. AP-$LuH_2$ exhibits a shiny blue color at ambient conditions with the typical particle size of 40-50 μm, as shown in the inset of Figure 1(b). Powder XRD patterns in Figure 1(b) confirm that the cubic $LuH_2$, with a fluorite structure (space group of $Fm$-$3m$) and lattice parameter, a = 5.0319 Å, is the major phase, and it coexists with a minor impurity phase of the Lu metal. For comparison, Figure 1(c) shows the XRD pattern and Rietveld refinement of the nitrogen-doped $LuH_{2-x}N_y$ (HP-$LuH_2$) sample. The experimental data can also be well fitted with the space group of $Fm$-$3m$ with a slightly smaller lattice parameter, a = 5.0310 Å, and a small amount of raw LuN reactant can be identified.

The energy dispersive X-ray spectroscopy (EDS) is applied to analyze the element composition in the samples. The Lu element and small amounts of C and O elements were identified in both AP-$LuH_2$ and HP-$LuH_2$ samples. However, note that the atomic number of H is too small to be detected using this technique. Furthermore, no evident nitrogen element was detected in the HP-$LuH_2$ sample, suggesting that the amount of N element that have been doped in $LuH_{2-x}N_y$ compounds is very little. The compositions were further investigated using the electron energy-loss spectroscopy (EELS) in TEM. As shown in Figure 2(a) and (b), the spectra were measured within two energy ranges (300-800 and 1450-1960 eV), The peak above ~1600 eV displays the $M_{4,5}$ edge of Lu. Note that no evident peak corresponding to nitrogen can be identified around 400 eV.

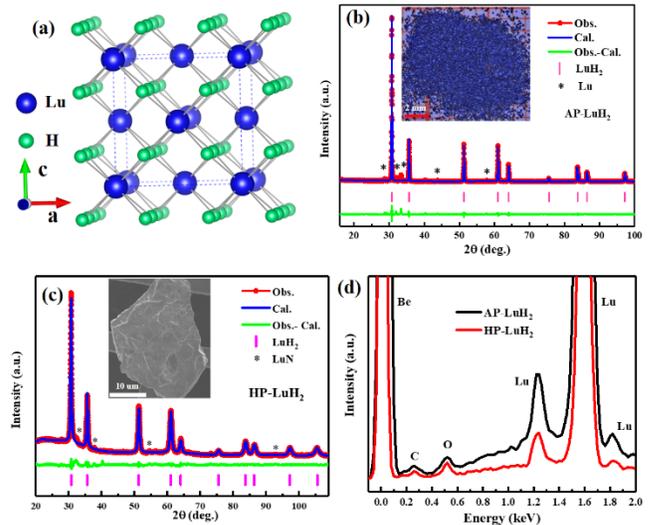

**Figure 1.** Structural model and powder X-ray diffraction data. (a) Structural model of AP-$LuH_2$ with the fluorite structure. The blue and green spheres represent lutecium and oxygen atoms, respectively. (b) PXRD patterns of the AP-$LuH_2$ and Rietveld fitting curve (blue line) of the data. The asterisks indicate small amounts of Lu metal impurities. The inset shows the pictures of AP-$LuH_2$ samples, exhibiting a dark-blue color. (c) Powder XRD patterns of the as-synthesized HP-$LuH_2$ with small amounts of LuN impurities. The inset shows the photographs of the HP-$LuH_2$ crystal. (d) EDS patterns of AP-$LuH_2$ and HP-$LuH_2$, Lu element and small amounts of C and O elements were identified in both AP-$LuH_2$ and HP-$LuH_2$ samples.

$LuH_2$ and $LuH_3$ adopt the same space group, $Fm$-$3m$, with lattice parameter $a$ = 5.0330 Å [30] and 5.1629 Å[24], respectively, and the bulk $LuH_3$ is unstable at ambient conditions. In this study, HP-$LuH_2$ was synthesized at 10 GPa and 600°C. Therefore, we must verify that the HP-$LuH_2$ is with a $LuH_2$ structure instead of $LuH_3$. First, the Rietveld refinements on the powder XRD pattern confirmed that the lattice parameter of the as-prepared HP-$LuH_2$ was approximately, $a$ = 5.0310 Å, which is very close to $a$ = 5.0319 Å in AP-$LuH_2$. To further determine the chemical composition and the form of the corresponding elements in AP-$LuH_2$ and nitrogen-doped HP-$LuH_2$ compounds, X-ray photo-electron spectroscopy (XPS) spectra were collected. Figure 2(c) and (d) display the Lu $4f$ high-resolution spectra of the AP-$LuH_2$ and nitrogen-doped HP-$LuH_2$ samples. The Lu $4f$ curve was deconvoluted into three Gaussian fitting peaks. The energies around 7.97 eV and 9.37 eV are attributed to the $Lu^{2+}$ $4f_{7/2}$ (blue line) and $Lu^{2+}$ $4f_{5/2}$ (cyan line) in the AP-$LuH_2$ sample. These values are obviously different from those of $Lu^{3+}$ compounds. For the $Lu_2O_3$ compound, the $Lu^{3+}$ $4f_{7/2}$ and $Lu^{3+}$ $4f_{5/2}$ peaks are located at 8.3 eV and 9.8 eV, respectively. The corresponding energy values show conformance with $Lu^{2+}$ in the LuO compound[31]. The HP-LuH$_2$ sample displayed almost the same binding energies, the Lu $4f$ characteristic peaks $4f_{7/2}$ and $4f_{5/2}$ at 7.99 eV and 9.41 eV, respectively. The obtained results suggest that the Lu valence states are very close to 2+ on both AP-LuH$_2$ and HP-LuH$_2$, and the two samples have very similar chemical compositions.

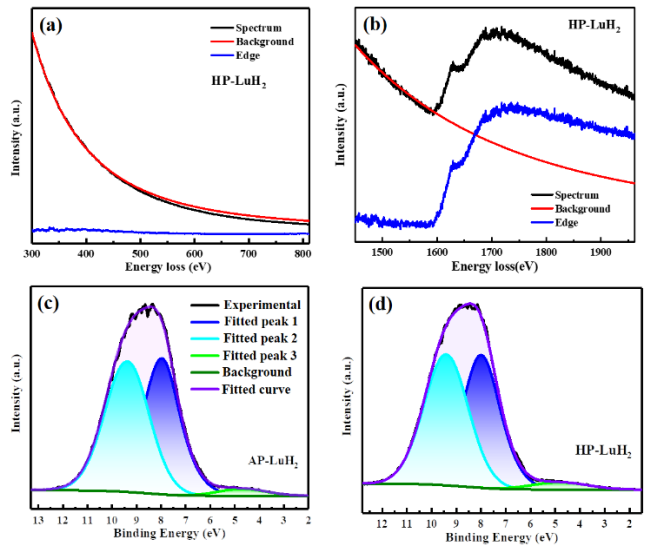

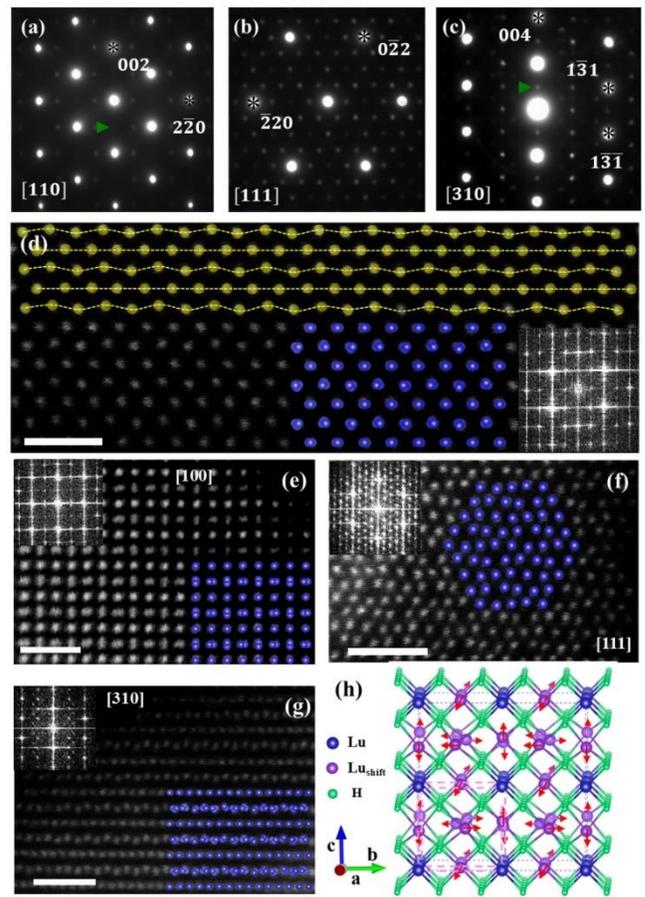

**Figure 2.** Characterization of the chemical composition for the AP-LuH$_2$ and HP-LuH$_2$ examples. (a, b) EELS spectra of HP-LuH$_2$ within two energy ranges (300–800 and 1450–1960 eV). No peak that typically indicates nitrogen was detected at 400 eV. The identified signal above ~1600 eV is the EELS spectra from Lu elements. Lu $4f$ XPS spectra for (c) AP-LuH$_2$ and (d) HP-LuH$_2$ examples. Fitting curves (solid curve) and Lu$^{2+}$ $4f_{7/2}$ (blue line), Lu$^{2+}$ $4f_{5/2}$ (green line), and O $2p$ peaks (green) are demonstrated.

Selected-area electron diffraction (SAED) and aberration-corrected STEM were used to conduct microstructural investigations on both AP-LuH$_2$ and HP-LuH$_2$ samples. Figure 3(a)-(c) show the electron diffraction patterns of LuH$_2$ taken along the [110], [310], and [111] zone axis directions at room temperature, respectively. The main diffraction spots with relatively strong intensities in the diffraction patterns can be well indexed with respect to the space group, $Fm$-$3m$, with $a$ = 5.01 Å, this result is consistent with the result obtained through the XRD measurement. On the other hand, the most striking structural phenomenon revealed in our TEM observations is the appearance of a series of superlattice spots following with the main diffraction spots, can be characterized by a modulation wave vector of $\boldsymbol{q}^* = \frac{1}{4}(2\bar{2}0)$.

To better understand the atomic structural features and explore the origin of superlattice spots, we considered the atomic-resolution STEM images. The STEM images taken along the [110], [100], [111], and [310] directions are shown in Figure 3(d–g) with the corresponding fast-Fourier transform (FFT) images. The FFT patterns show similar superlattice spots as those observed in the SAED patterns. In addition, a thorough analysis of the STEM images showed a periodic shift of the Lu atoms. Figure 3(d) shows a typical STEM image taken along the [110] direction, clearly exhibiting the atomic displacement in the Lu layer. The displacements of the two types of Lu atom columns, Lu$_{down}$ (antiparallel to the $c$-axis) and Lu$_{up}$ (parallel to the $c$-axis), are evident. Note that only one of every two adjacent layers of the Lu atoms shift relative to each other, whereas all the atoms in the other layer remain in their original positions. Atomic displacements were also observed in other three directions, confirming that they are common in the AP-LuH$_2$ material. the aberration-corrected STEM images of AP-LuH$_2$ allowed us to directly measure the shift in the Lu atoms relative to the original positions, estimated as 40 pm. Based on the experimental data, a new $2 \times 2 \times 2$ supercell was constructed that included the ordered displacement of Lu atoms, as shown in Figure 3(h) (Detailed see Figure. S1, Supporting Information). The blue and violet spheres represent the original and shifted Lu atoms, respectively. According to the new atomic model, the simulated diffraction patterns well reproduce the experimental results, as shown in Figure.S2. In addition, *in-situ* electron diffraction experiments on AP-LuH$_2$ also demonstrated that the superstructure is stable over a broad temperature range of 94-673 K as shown in Figure S3.

**Figure 3.** (a-c) Selected-area electron diffraction (SAED) patterns of AP-LuH$_2$ taken along the [110], [111], and [310] zone axis directions. A modulated wave vector, $\boldsymbol{q}^* = \frac{1}{4}(220)$, can be confirmed along the (a) [110] and (b) [111] directions, respectively. (d-g) High-angle annular dark field (HAADF) images recorded along the [110], [100], [111], and [310] zone axis directions with the corresponding fast-Fourier transform (FFT) images. (h) A structural model with Lu atoms displacement was constructed based on the experimental observations, and the corresponding structural models were superimposed on the HAADF images. Scale bar is 1 nm.

We also performed systematic microstructure investigations of the nitrogen-doped HP-LuH$_2$ compound. The modulated crystal structure can be preliminarily reflected by the SAED patterns along the [110], [100] and [310] zone axis directions, as shown in Figure 4(a)-(c), and further supported by the HAADF images in Figure 4(d)-(e), which are highly consistent with those of AP-LuH$_2$. It is remarkable to note that antiphase in Figure 4(d) and (e). Such a kind of atomic structural features are like the domain walls observed in hexagonal manganites, RMnO$_3$(R = Ho-Lu, Y, and Sc), which are usually closely related to many intriguing physical properties, such as multiferroic and interface transport[32-35].

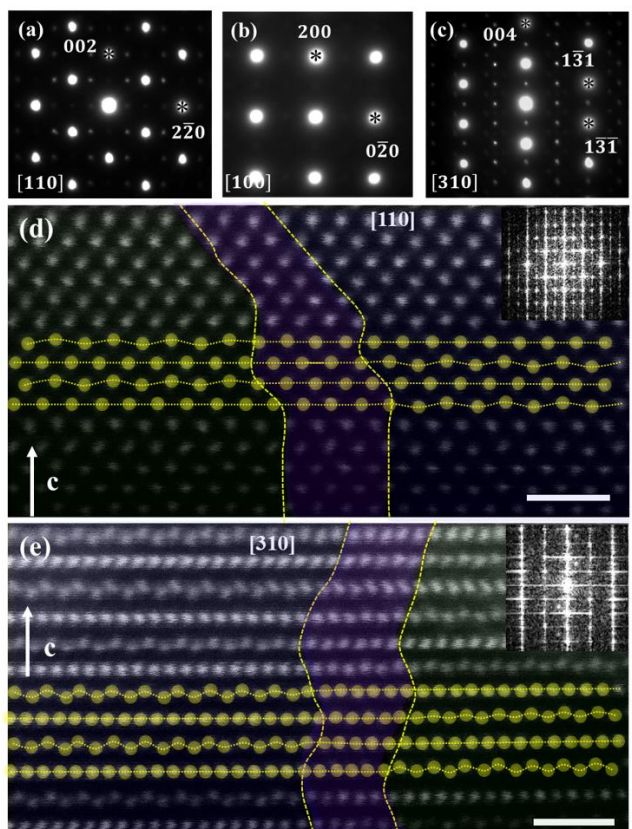

**Figure 4.** (a-c) SAED patterns of HP-LuH$_2$ taken along the [110], [111], and [310] zone axis directions. (d-e) HAADF images reco domain boundaries with transition region of a few atomic layers can be frequently observed in the HP-LuH$_2$ materials, as shown rded along the [110] and [310] zone axis directions with the corresponding FFT images inserted, respectively. An antiphase boundary, crossing a domain wall as indicated by the red region, can be directly obtained in HAADF images taken along [110] and [310] zone axis directions. Scale bar is 1 nm.

The electronic and structural evolutions of a material are often codependent and evolve synchronously. Our structural investigations in both AP-LuH$_2$ and HP-LuH$_2$ materials demonstrated the presence of ordered Lu atom displacement. The existence of remarkable Lu atoms displacement will inevitably change their chemical environments, such as the distortion of hydrogen octahedron and even the introduction of hydrogen vacancies, which further affect the electronic properties. Clearly, more experimental techniques, especially for light elemental sensitivity experiments, are needed to determine the exact crystal structure and stoichiometry of nitrogen-doped lutetium hydride. A better detailed structural description will enable the theoretical modeling of these nonstoichiometric metal hydrides and improved theoretical understanding.

## ■ CONCLUSION

In summary, we investigated the microstructural properties of AP-LuH$_2$ and HP-LuH$_2$ by using atomically resolved STEM techniques. A clear superstructure with a modulation wave vector of $\boldsymbol{q}^* = \frac{1}{4}(2\bar{2}0)$ due to Lu atoms displacement was observed in both samples and antiphase domain boundaries with transition region of a few atomic layers can be frequently observed in the HP-LuH$_2$ materials, suggesting some reported interesting theoretical and experimental results might correlated with these micro-structural features.

## ASSOCIATED CONTENT

**Supporting Information**.
Detailed experimental procedures, *In-situ* electron diffraction results, and TEM simulation.

## AUTHOR INFORMATION

**Corresponding Author**
hxyang@iphy.ac.cn
ljq@aphy.iphy.ac.cn
jgcheng@iphy.ac.cn

**Notes**
The authors declare no competing financial interest.

## ACKNOWLEDGMENT

This work was supported by the National Natural Science Foundation of China (Grant Nos. U22A6005，12074408, 52271195, 12074425, 11874422, 12025408,11921004), the National Key Research and Development Program of China (Grant Nos. 2021YFA13011502, 2019YFA0308602, 2021YFA1400200), the Strategic Priority Research Program (B) of the Chinese Academy of Sciences (Grant Nos. XDB25000000, XDB33000000), the Scientific Instrument Developing Project of the Chinese Academy of Sciences (Grant Nos. YJKYYQ20200055, ZDKYYQ2017002, 22017BA10), the Synergetic Extreme Condition User Facility (SECUF), the Guangdong Major Scientific Research Project (2018KZDXM061), Beijing Municipal Science and Technology major project(Z201100001820006) and IOP Hundred Talents Program (Y9K5051).

# Supporting Information

# Microstructure and structural modulation of lutetium dihydride $LuH_2$ as seen via transmission electron microscopy

Xiao-Ping Ma,[1, 2, 3, ‡] Ning-Ning Wang,[1, 5, ‡] Wen-Tao Wang,[1, 5] Jing-Zhe Nie,[1, 5] Wen-Li Gao,[1] Shuai-Shuai Sun,[1, 5] Jun Li,[1, 5] Huan-Fang Tian,[1, 5] Tian-Long Xia,[2, 3, 4, 6] Jin-Guang Cheng,[1, 5, *] Jian-Qi Li,[1, 5, 7, *] and Huai-Xin Yang[1, 5, *]

1 Beijing National Laboratory for Condensed Matter Physics and Institute of Physics, Chinese Academy of Sciences, Beijing 100190, China

2 Department of Physics, Renmin University of China, Beijing 100872, China

3 Beijing Key Laboratory of Opto-electronic Functional Materials & Micro-nano Devices, Renmin University of China, Beijing 100872, China

4 Key Laboratory of Quantum State Construction and Manipulation (Ministry of Education), Renmin University of China, Beijing, 100872, China

5 School of Physical Sciences, University of Chinese Academy of Sciences, Beijing 100190, China

6 Laboratory for Neutron Scattering, Renmin University of China, Beijing 100872, China Yangtze River Delta

7 Songshan Lake Materials Laboratory, Dongguan, Guangdong, 523808, People's Republic of China

**Experimental procedures**

In this study, we used the commercially available $LuH_2$ powder (JiangXi Viilaa Metal Material Co., Ltd.) as AP-$LuH_2$ sample as described in a previous paper[36]. The nitrogen doped HP-$LuH_2$ samples were prepared by sintering a mixture of AP-$LuH_2$ and LuN powders in the ratio of 95:5 under high-pressure and high-temperature (HPHT) conditions in a Kawai-type multianvil module (Max Voggenreiter GmbH). The mixture was contained in a sealed Pt capsule, inserted into an h-BN sleeve, and then placed in a graphite heater. All these sample assembly was put in the central hole of a semi-sintered MgO ceramic octahedron that served as the pressure transmitting medium. During the HPHT experiments, the pressure was first increased to 10 GPa, and then the temperature was increased to 600 °C and kept for 1 h before quenching to room temperature. The sample was recovered and characterized at ambient pressure.

Powder X-ray diffraction (XRD) measurements were carried out with a Bruker AXS D8 Advance diffractometer equipped with Cu $K\alpha$ radiation at 40 kV and 40 mA with 2θ ranging from 10° to 110°, a step with of 0.02°and a counting time of 0.5 s/step. The Rietveld refinements were done by using the software of Jana2020. The chemical composition and microstructure analysis were performed on a Hitachi model S-4800 field emission scanning electron microscope (SEM) with an energy-dispersive spectrometer (EDS). Atomic-resolution HAADF images were obtained with a spherical-aberration-corrected JEOL ARM200F instrument equipped with double-aberration correctors and operated at 200 kV. The transmission electron microscope (TEM) samples were prepared by crushing the polycrystalline into fine fragments, then the resulting suspensions were dispersed on holey copper grids.

*In-situ* electron diffraction experiments on AP-$LuH_2$ demonstrated that the superstructure is stable with respect to temperature. Figure S1 shows a series of SAED patterns taken at different temperatures, showing that the superstructure spots persist at 94–673 K. We extracted the normalized diffraction intensity and distance between the main diffraction and satellite spots. Figure S1(h) exhibits the evolution of diffraction-spot intensity with temperature. No obvious changes were observed in the diffraction intensity from 94 to 673 K. The distance between the diffraction spots also remains constant, as shown in Figure S1(l). Therefore, the *in-situ* experimental data suggested that the LuH$_2$ superstructure was stable over a broad temperature range of 94–673 K.

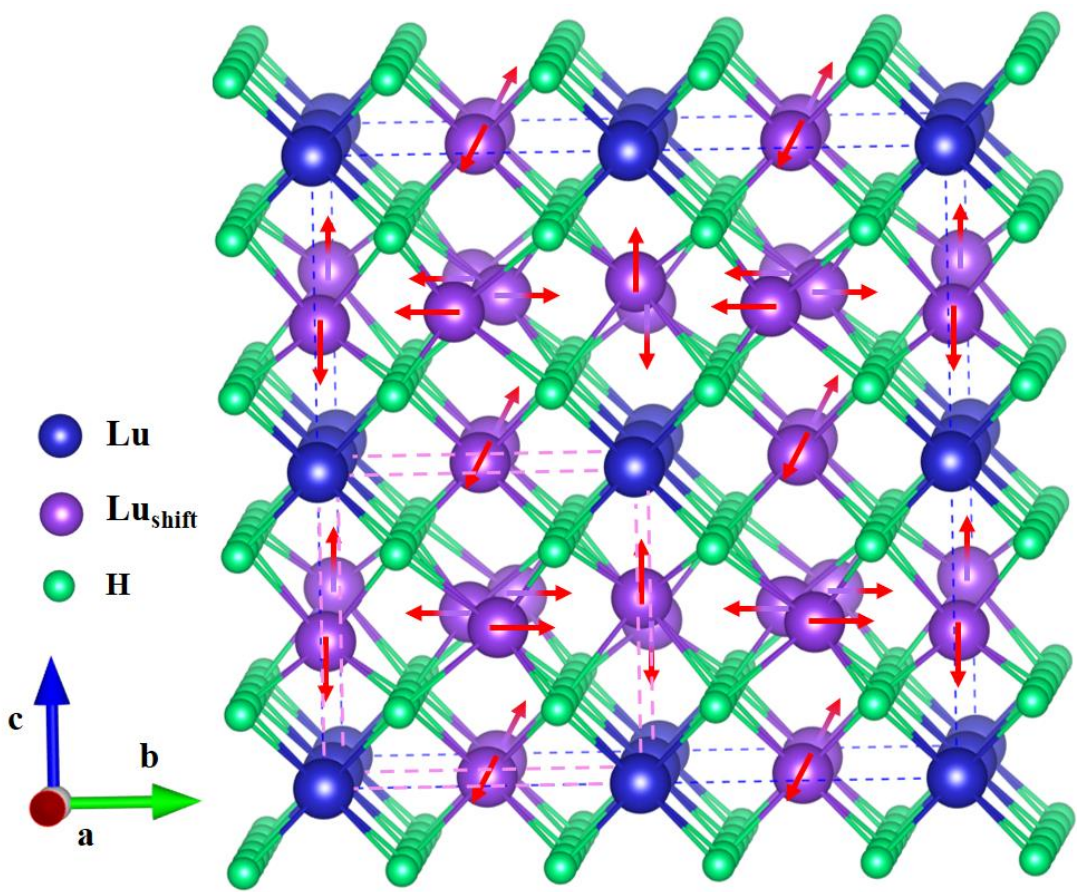

**Figure S1.** A new 2× 2× 2 supercell was constructed including the ordered displacement of Lu atoms according to STEM data. The blue spheres, representing the Lu atoms at the vertex of the fluorite type LuH$_2$ crystal structure, does not shift in the new superlattice. Then, the purple spheres, at face center sites of the original LuH$_2$ crystal structure, relative shift to each other only along the a, *b*, or *c* zone axis. The corresponding new structural models were superimposed on the HAADF images as shown in **Figure 3**. Additionally, it is noted that the hydrogen is too small to be detected using STEM technique, we cannot obtain information about hydrogen atoms, and the hydrogen atoms in the supercell are the same as in the fluorite type LuH$_2$. The outer blue dotted rectangle represents the unit cell of new atomic model, and the pink dotted rectangle represents the unit cell of the original fluorite type LuH$_2$ crystal structure.

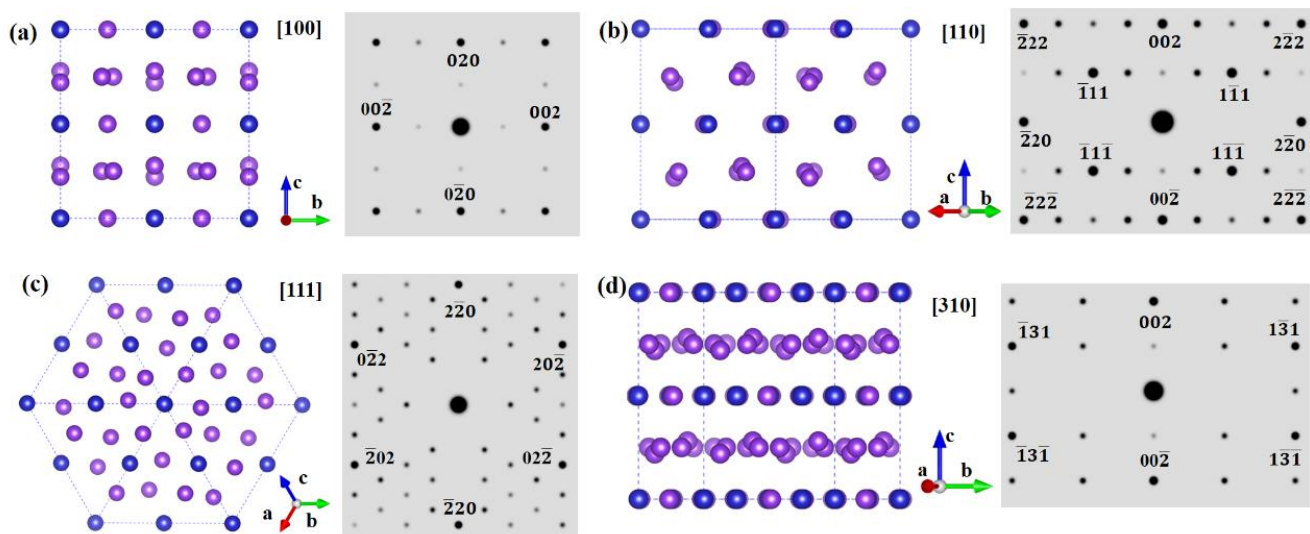

**Figure S2.** The simulated diffraction patterns along different zone axis based on the supercell crystal structure model as shown in **Figure S1.** (a-d) The simulated diffraction patterns are consistent with experimental diffraction patterns as shown in **Figure 3**.

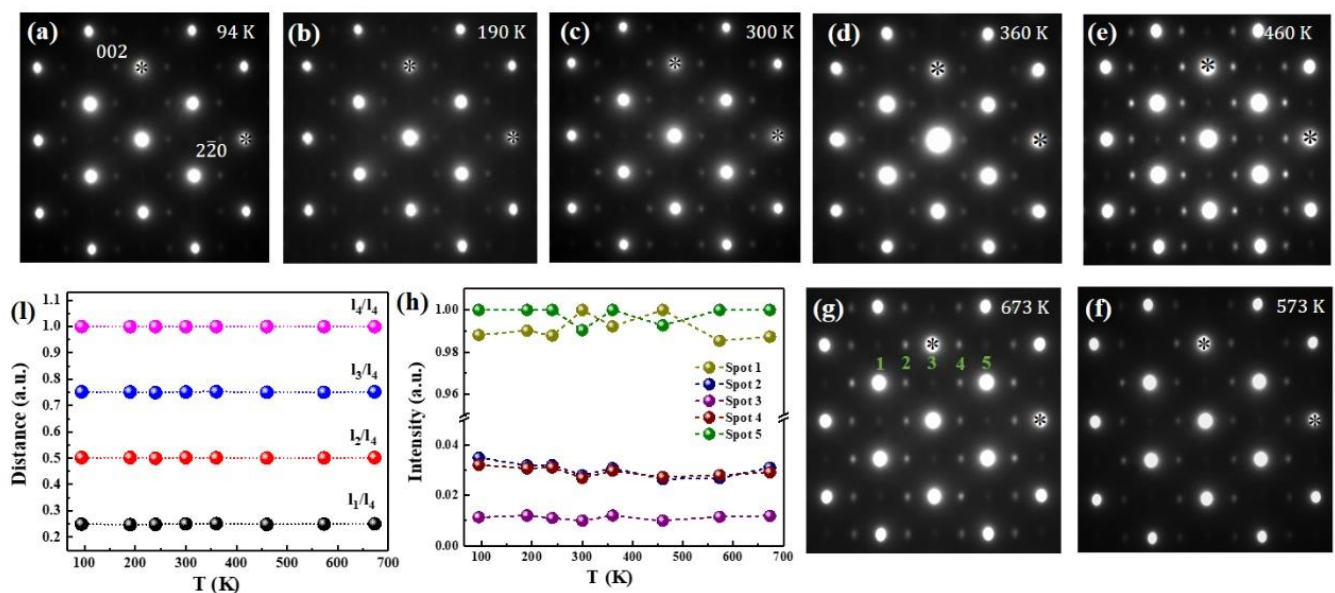

**Figure S3.** (a–g) A series of selected-area electron diffraction (SAED) patterns of AP-LuH$_2$ along the [110] zone axis direction at temperatures of 94–673 K. (h) Normalized intensity of the selected diffraction spots (I$_1$: the diffraction intensity of spot 1, I$_2$: the diffraction intensity of spot 2, I$_3$: the diffraction intensity of spot 3, I$_4$: the diffraction intensity of spot 4, I$_5$: the diffraction intensity of spot 5) as a function of temperature. (l) Evolution of the normalized distance (l$_i$/l$_4$) between the selected diffraction spots (l$_1$: the distance between spots 1 and 2, l$_2$: the distance between spots 1 and 3, l$_3$: the distance between spots 1 and 4, l$_4$: the distance between spots 1 and 5) with temperature.